\documentclass[12pt]{arXiv_class}

\title{\textbf{An Optimal Polarization Tracking Algorithm for Lithium-Niobate-based Polarization Controllers}} 
\author{Joaquim D. Garcia and Gustavo C. Amaral}

\begin{document}
\maketitle

\begin{abstract}
We present an optimal algorithm for the three-stage arbitrary polarization tracking using Lithium-Niobate-based Polarization Controllers: device calibration, polarization state rotation, and stabilization. The theoretical model representing the lithium-niobate-based polarization controller is derived and the methodology is successfully applied. Results are numerically simulated in the MATLAB environment.
\end{abstract}

\section{Introduction}

Keeping the polarization state stable in long-haul fiber optical communication links is a difficult task due to the local changes in the silica structure along the fiber which induces bi-refringence and, therefore, variation of the polarization state \cite{AgrawalBOOK2002,DericksonBOOK1998}. Although until the early 1990's most optical communication links did not account for it, Polarization Mode Dispersion (PMD) has been a more serious threat to modern optical links as the bit rate grows \cite{GisinJLT1991}. In this context, polarization control has gained much attention, specially with the development PMD-compensation techniques \cite{gordon2000pmd} and of the so-called Polarization Shift Keying \cite{benedetto1992theory}.

Lithium-niobate ($LiNbO_3$) is a material capable of altering its refractive index upon application of a difference of potential between its terminals \cite{van1993modeling}. This device represented a huge step in the polarization stabilization and control technology since it allowed extremely fast polarization controlling and tracking devices to be developed, once no mechanical structures were necessary \cite{koch2008optical}. In this work, we present a complete algorithm for polarization control and stabilization that relies on the use of the aforementioned $LiNbO_3$ structures, more specifically, the EOSpace Polarization Controller Module \cite{EOSpace}. The polarization control itself is composed of two main steps: firstly, an analytical rotation along the Poincar\'e Sphere relying on basic analytic geometry \cite{stein1977calculus} and quaternion arithmetic \cite{kuipers1999quaternions}; secondly, stabilization method based on adaptive filtering to achieve fine adjustment \cite{haykin1999adaptive}. A state estimator that is fundamental for the good functioning of the rotation algorithm is also presented.

\section{Mathematical representation of polarization}\label{mathRep}

The state of polarization of light has been represented mathematically as Jones Vectors and Stokes Vectors \cite{saleh1991fundamentals}. We shall stick to the Stokes representation since visualization in the Poincar\'e Sphere is direct. Neverthelees, the conversion between these two representations require straightforward computations. Stokes vectors are 4-dimensional vectors that carry information about the State of Polarization (SOP) of light. Since the first component ($S_0$) is associated to the total light intensity, it is common to normalize the Stokes Vector by dividing it by $S_0$. In the case of coherent light, the 3-dimensional Stokes Vector formed of the remaining three normalized components of the former 4-dimensional vector, has norm $1$. Since we have 3-dimensional normalized vectors representing the SOP, it is usual to represent it graphically in a 3-sphere known as the Poincar\'e sphere. Since all SOPs are mapped bijectively in the 3-sphere, we shall treat, from now on, an SOP as a point in the 3-sphere.

SOP changes that do not affect the light intensity can be represented by rotations in 3-space. These rotations are a class of unitary transformations and can be represented by orthonormal matrices \cite{strang09}. Even though, polarisers do affect the light intensity and, as such, cannot be represented by orthonormal matrices, they can be represented by projection matrices. The rotation matrix in 3-space has a very interesting characterization via the Spectral Theorem: they always have 3 eigenvalues (since they are normal); all of the eigenvalues have norm 1; one of the eigenvalues is always equal to 1 and its eigenvector is the rotation axis, $e$; the remaining eigenvalues are complex conjugate numbers whose real part correspond to the cosine of the rotation angle, $\theta$, and whose imaginary part correspond to the sine of the rotation angle.

The quaternions are a number system that extend the complex numbers. Since the unitary quaternions are homeomorphically mapped to the 3-space rotation matrices \cite{crossley2006essential}, it is possible to use them to perform 3-space rotations for which they were shown to be way more stable \cite{karlsson2004quaternion}. For the aforementioned reasons, our algorithm will rely on quaternions representation instead of on the matrix representation.

\section{Lithium-Niobate-based Polarization Controller Characteristics}\label{Control}

Literature around polarization control is very rich and diverse techniques were proposed and verified along the last years: \cite{heismann1994analysis69} shows that three elements, two quarter wave plates and one half wave plate, are sufficient to reach any SOP from any other SOP; other methods appear in \cite{imai1985optical46,noe1988automatic40,heismann1989integrated55}. As mentioned before our methodology focuses on the electro-optic $LiNbO_3$ EOSpace Polarization Controller Module (PCM). Such device is available commercially as a multi-stage component but, for simplicity, we are going to develop an algorithm that uses a single stage. The algorithm is easily extended to account for the multi-stage version to reduce the input voltage that may vary within a $\pm70$ Volts range. 

A single stage of the PCM has 3 electrodes \cite{EOSpace} and realizes an arbitrary \textit{Linear Retarder}: linear wave plates that induce relative phase differences between the TE and TM modes of the propagating electromagnetic field due to the variation of the refractive index in both axes as a function of the applied difference of potential which causes birefringence and, thus, altering the polarization state. Linear Retarders have a main polarization axis, also known as eigen-mode, $e\in \{v\in\mathbb{R}^3| z=0\}$, and a characteristic phase delay, $\theta\in[0, 2\pi)$. It is possible to show that, by changing the eigen-mode and the phase delay of a linear retarder, one can shift from one SOP to any other SOP. The proof of this is actually constructive and our algorithm provides such construction. In order to  set the eigen-mode to $e=(cos(\alpha/2), sin(\alpha/2),0)$ and the phase delay to $\theta = 2\pi\delta$ the electrodes voltages must be set to:
\begin{align}
&V_a=2V_0\delta sin (\alpha) -V_{\pi}\delta cos(\alpha)+V^b_{a}\label{ds_EOS1} \\
&V_b=0 \label{ds_EOS2} \\
&V_c=2V_0\delta sin (\alpha) -V_{\pi}\delta cos(\alpha)+V^b_{c}\label{ds_EOS3} 
\end{align}
where $V_{\pi}$ is the voltage required to induce a $180^o$ phase shift between the TE and TM modes for a single stage, $V_0$ is the voltage required to rotate all power from the TE to the TM mode, or vice versa, for a single stage, and $V^b_{a}$ and $V^b_{c}$ are the bias voltages required on electrodes A and C, respectively, in order to achieve zero birefringence between the TE and TM modes \cite{EOSpace}. Even though the data-sheet of the device provides the range within which $V_0$, $V_{\pi}$,$V^b_{a}$ and $V^b_{c}$ should be, their actual values for an arbitrary stage must be determined via a calibration procedure. The calibration procedure adopted in our methodology is presented in \cite{xi2010novel}.

\section{Analytical Rotation Algorithm}\label{rotAlg}

The feasible eigen-modes of Linear retarders are linear SOPs, hence their name, so the rotation axis must lie in the $s_1$-$s_2$ plane, i.e., along the equator of the Poincar\'e Sphere. Therefore, given two polarization states represented by $s_{in}$ and $s_{target}$, we must find a rotation axis lying on the $s_1$-$s_2$ plane and a rotation angle such that the corresponding linear transformation converts $s_{in}$ into $s_{target}$. Since the possibility of measuring the output SOP after a transformation is of paramount importance of our adaptive algorithm, we assume, throughout the development, that a polarimeter such as the one described in \cite{CalliariTCC2014} is included in the control loop.

\begin{figure}[ht]
\centering
\includegraphics[width=0.9\linewidth]{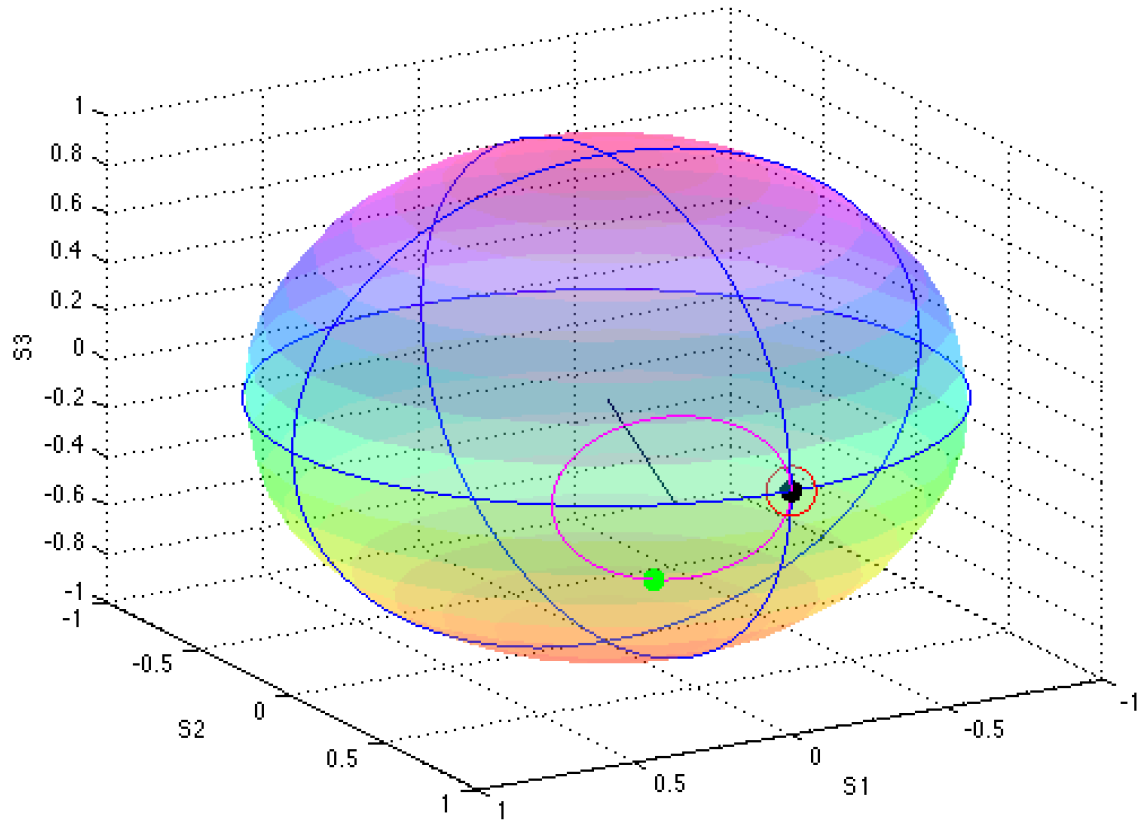}
\caption{Rotation example: input SOP in green, target SOP in black, output SOP in red, rotation axis in black, rotation circle in pink. The rotation arc is the smaller arc (least angular distance) defined by the points in the pink circle.}
\label{fig:find_rot}
\end{figure}

We start the rotation step by defining the following vectors: $v_1$, orthogonal to the rotation axis; $v_2$, the normalized cross-product between $v_1$ and $s_3$ (supposing $v_1$ is not parallel to $s_3$); $v_3$, the centre of the rotation, i.e., the point in the rotation plane that intersects the line parallel to the rotation axis. In the case that $v_1$ is parallel to $s_3$, we use the first of these that is not parallel to $v_1$: $s_{in}$, $s_{target}$ or $s_1$. Now, we set: $v_4=s_{in}-v_3$; and $v_5=s_{target}-v_3$ where one can imagine $v_4$ and $v_5$ as two clock arrows where the rotation angle, $\theta$, is the oriented angle between them. The cosine of $\theta$ is given by:
\begin{align}
cos(\theta)=\frac{\langle v_4,v_5\rangle}{||v_4||_2||v_5||_2}
\end{align}
Note that the angle $-\theta$ has the same cosine as $\theta$. To determine which rotation direction is the right one, we take the sign of the cross product between $v_4$ and $v_5$ with the following implication: if its negative, then $\theta=-\theta$ and $v_3=-v_3$; if its positive, then do nothing. After all those computations, $\theta$ is the rotation angle and $v_3$ is the rotation axis. Fig. \ref{fig:find_rot} presents the graphical interpretation on the 3-space Poincar\'e Sphere of the determination of both rotation axis and rotation angle given two arbitrary SOPs for $s_{in}$ and $s_{target}$.

\section{Stabilization algorithm}\label{sgdAlg}

Generally, non-linear optimization problems such as the one presented suffer from an intrinsic issue: finding a rotation that changes from one polarization state to another is extremely useful when the SOPs are distant but become highly unstable when $s_{in}$ and $s_{target}$ are close to each other. Taking the Levenberg-Marquadt procedure as inspiration \cite{}, we devise a two-step control algorithm that alternates between rotation and stabilization depending on the distance between SOPs. The stabilization step also helps eliminating problems involving measurement errors and numerical approximations that may influence the stability. For this step, we resort to an old optimization algorithm known as the \textit{Gradient Descent}, which has been successfully applied to adaptive filtering \cite{widrow1976stationary} and machine learning \cite{burges2005learning}.

Given a cost function $J(x)$ to be minimized, where $x$ is a $n$-dimensional vector, the Gradient Descent algorithm updates $x$ in the direction of the negative (descending) gradient, in search for the minimum of the functional, with the rule $x=x-\alpha \nabla J(x)$, where $\alpha$ is a step size parameter that must be calibrated \cite{haykin1996adaptive}. Since the cost function depends non-linearly on the input and target SOP's, there is no simple analytical form for the gradient and it must be estimated by measurements with intrinsic error. This measurement error is the reason why we employ a variant of the algorithm known as the Stochastic Gradient Descent (SGD) \cite{shynk2012probability}.

It is possible to estimate all the components of the gradient $\nabla J(x)$ by using the well-known secant method. Given $e_i$, the $i^{th}$ element of the canonical basis of $\mathbb{R}^n$, we can perturb the current value $x$ by a small value $\epsilon$, and accounting for measurement imprecisions by evaluating the cost function $m$ times for a better estimate, the $i^{th}$ component of $\nabla \hat{J}(x)$ is approximated by:
\begin{align}
\nabla \hat{J}(x)_{i} \approx \frac{1}{m}\sum^m_{k=1}\frac{J^k(x+\epsilon e_i)-J^k(x-\epsilon e_i)}{\epsilon}
\end{align} 

When the gradient is estimated instead of being computed analytically, the descending path is way less smoother, and it is usual to see some roaming around the local minimum \cite{}. For our problem, $x=[ V_a , V_c ]^\top$ and $J(x) = || S_{out}(x) - S_{in} ||^2_2$ since the 3-Space output Stokes Vector is a function of the linear retarder input voltages. The algorithm was simulated numerically via MATLAB with the step $\alpha$ empirically set as directly proportional to the error between $s_{target}$ and $s_{in}$.

\section{State estimation}\label{seAlg}

For a given calibration, we have estimates of the values $V_0$, $V_{\pi}$,$V^b_{a}$ and $V^b_{c}$. Thus, at any time, if we know the voltages $V_a$ and $ V_c$ applied to the PCM stage's electrodes, we can obtain $\alpha$ and $\delta$ by solving a $2\times2$ non-linear system which has an analytical solution. With the pair $(\alpha,\delta)$, we can easily obtain the pair $(e,\theta)$ and figure out the rotation implemented by the stage and, by taking its inverse and applying on $S_{out}$, we can obtain an estimate of $S_{in}$. Let $A=2V_0$, $B=-V_{\pi}$, $C=V_a-V^b_a$, and $D=V_c-V^b_c$. Then:
\begin{align}
&\delta = \sqrt{\left(\left(C+D\right)/\left(2A\right)\right)^2+\left(\left(C-D\right)/\left(2B\right)\right)^2}\\
&2\alpha = f\left( \left(C+D\right)/\left(2A\delta\right),\left(C-D\right)/\left(2B\delta\right) \right)
\end{align}
where $f(\cdot,\cdot)$ is the function that receives $\cos(\alpha)$ and $\sin(\alpha)$ and returns $\alpha$. It is worth emphasizing that an analogous procedure can be performed if the calibration resorts on LUTs \cite{hidayat2008high}.

\begin{figure}[ht]
\centering
\includegraphics[width=0.9\linewidth]{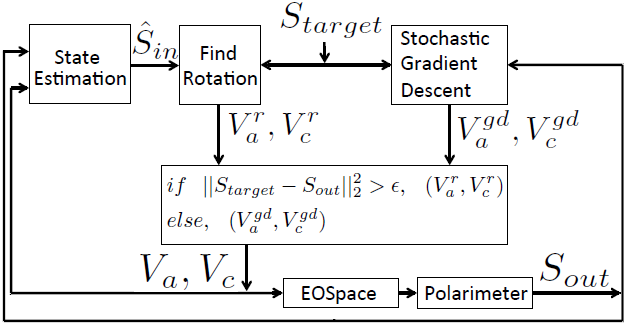}
\caption{Workflow chart diagram of the proposed three-stage control algorithm.}
\label{fig_alg}
\end{figure}

The complete two-stage methodology to control and to stabilize polarization making use of the rotation algorithm , the stochastic gradient descent algorithm and, the state estimator is presented in Fig.\ref{fig_alg} in the form of a workflow chart. It contains two main loops that compute the voltages to be applied to the Polarization Controller stage's electrodes. The choice for which pair of voltages to be used depends on the distance $\varepsilon$ between the output SOP and the target SOP: when $S_{out}$ and $S_{target}$ are distant, the SGD algorithm may take a long time to reach stability so the rotation algorithm is employed; on the other hand, if they are close enough, the rotation algorithm can be unstable so the SGD is employed.

\section{Simulation Results}

\begin{figure}[ht]
\centering
\includegraphics[width=0.9\linewidth]{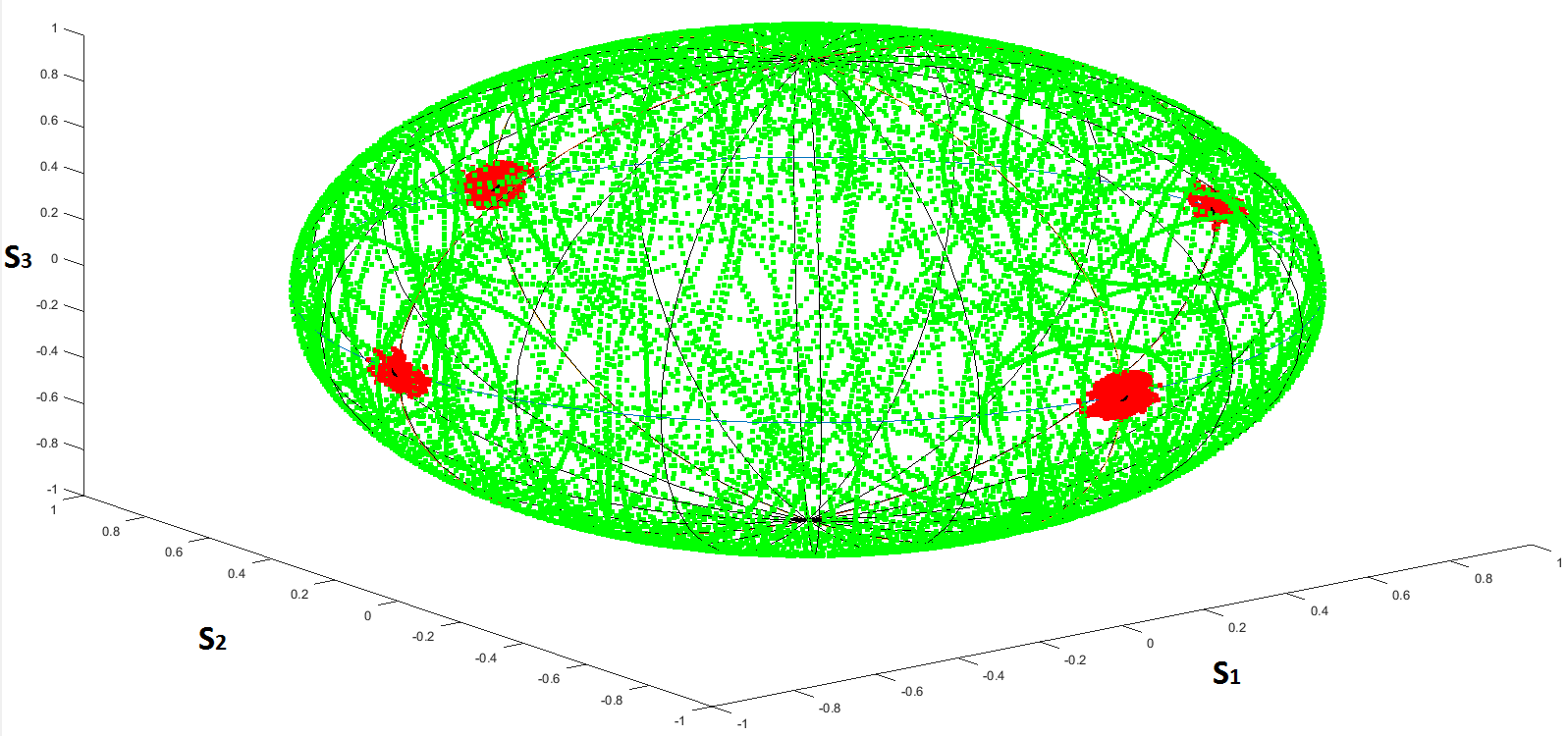}
\caption{Simulation of real-time polarization control considering both the inherent polarization drift in the output of the controller and the target polarization shift. The green dots represent the wandering state of polarization at the input of the controlling apparatus and the red dots represent the output state of polarization. The algorithm is capable of maintaining the polarization stable in the vicinity of $s_{target}$.}
\label{PoincareSphere_TrackingResult}
\end{figure}

In our simulation procedure, two issues were taken into account: the polarization drift of $s_{in}$, which is inherent to fiber optical communication systems; and the necessity of shift between two or more $s_{target}$'s. To clearly depict the algorithm's performance through the simulation results, we present, in Fig. \ref{PoincareSphere_TrackingResult}, the Poincar\'e Sphere and, in it, the green dots represent the wandering state of polarization at the input of the controlling apparatus and the red dots represent the output state of polarization. The algorithm is capable of maintaining the polarization stable in the vicinity of the set of $s_{target}$ defined by the three main linear states of polarization states: horizontal; vertical; diagonal; and anti-diagonal.

\begin{figure}[H]
\centering
\includegraphics[width=0.95\linewidth]{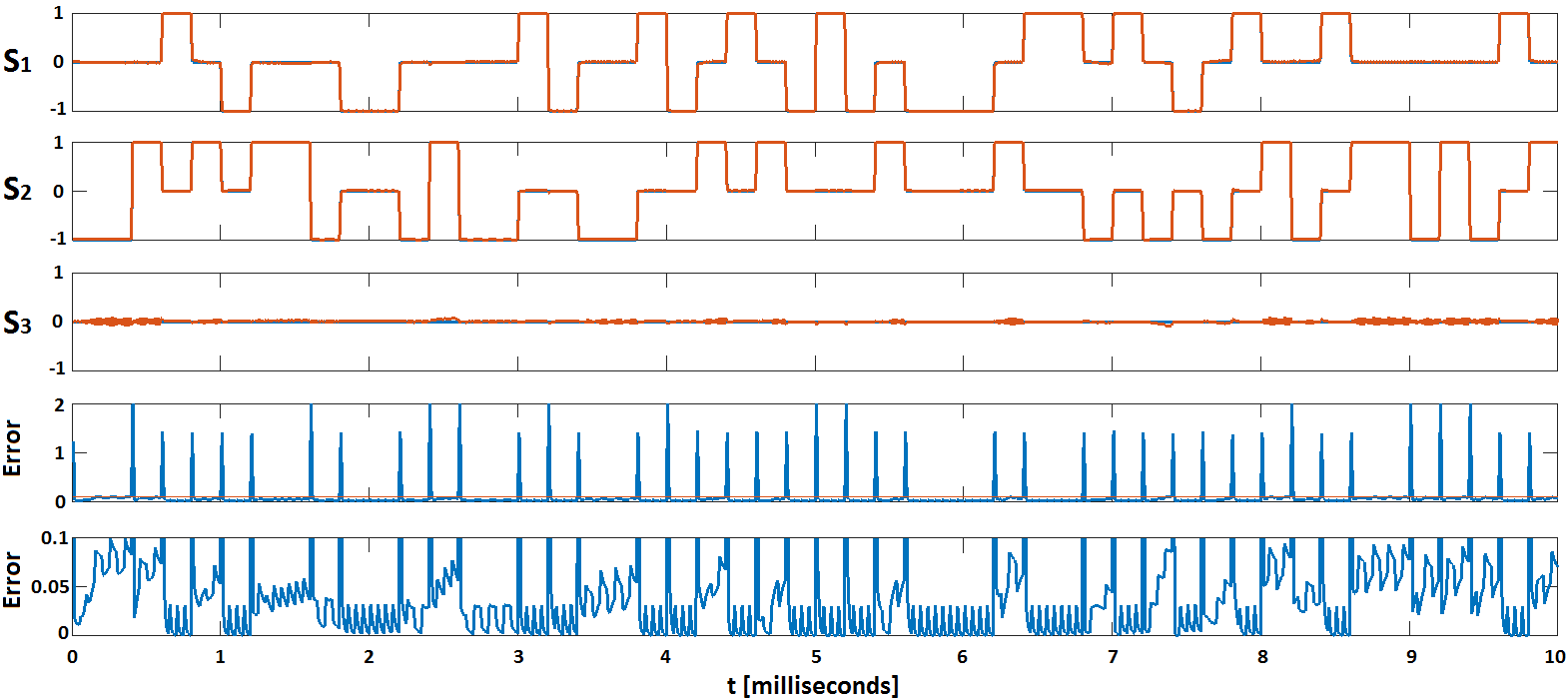}
\caption{Simulation of real-time polarization control considering both the inherent polarization drift in the output of the controller and the target polarization shift. All graphs are in the same horizontal scale displayed at the bottom of the figure. The first three graphs display the values of the Stokes parameters that compose the SOP vector. The bottom two graphs display the associated error between $s_{out}$ and $s_{target}$, where the second one is a re-scaled version of the first to clarify that the error does not exceed the threshold between SGD and the rotation step except in the cases of target polarization shift.}
\label{Parameters_TimeVar}
\end{figure}

In Fig. \ref{Parameters_TimeVar}, we present the values of each component of $s_{out}$ and $s_{target}$, as well as the associated error between $s_{target}$ and $s_{out}$, as a function of time. We observe that the algorithm uses the rotation step only after the shifts in $s_{target}$, while the SGD is responsible for stabilizing the polarization around its value while $s_{in}$ drifts. This result is in correspondence to the expected behaviour of the algorithm, and confirms its good performance and applicability since the associated error is very small, i.e., even though the vicinity into which the algorithm keeps the output state of polarization in relation to the target polarization state seems large when observing Fig. \ref{PoincareSphere_TrackingResult}, we see that, in the majority of time, inside a smaller vicinity define as the algorithm's error tolerance.

The time scale was determined based on the time responses of off-the-shelf Analog-to-Digital and Digital-to-Analog Converters (ADC and DAC, respectively), of General Photonic's Polarimeter module, and of the EOSpace Polarization Controller Module \cite{TI_ADC,GeneralPolaDetect,TI_DAC,EOSpace}. It yielded an overall control loop iteration of approximately $1$ $\mu$s. The polarization shift was set to match a Polarization Shift Keying system working at $50\times10^3$ symbols per second. The drift in polarization was set at $6$ krad/s, a value well that attempts to mimic the operation of an optical communication system under severe conditions.

\section{Conclusion}\label{conclusion}

We presented a three-step methodology to calibrate a Lithium-Niobate-based polarization controller module, to control the polarization and to stabilize the output SOP. Calibration, rotation algorithm, the Stochastic Gradient Descent and the state estimation algorithm were successfully tested in MATLAB simulations. The algorithms presented take simple forms and are readily embeddable in an FPGA or micro-controlled unit so we leave it as a future point of investigation.

\section*{Acknowledgment}

The authors would like to thank brazilian agency CNPq for financial support. The authors are indebted to F. Calliari and V. Lima for help with the polarimeter unit and with the electronic circuitry, specially the analog-to-digital and digital-to-analog converters.

\section*{Supplemental Material}

The authors provide digital supplemental material to accompany the article: a video file depicting the simulation run of the algorithm can be accessed in \cite{YouTubeVideo}; and the Matlab source files necessary for running the algorithm and the simulation are available at \cite{GitAlgorithm}.

\bibliographystyle{IEEEtran}
\bibliography{Ref}

\vfill

\end{document}